%% file: main.tex
\documentclass[journal]{IEEEtran}
\input{preamble.tex}
\title{{Performance Analysis of a 5G Transceiver Implementation for Remote Areas Scenarios}}

\author
{
\IEEEauthorblockN{Wheberth~Dias, Danilo~Gaspar, Luciano~Mendes, Marwa~Chafii, Maximilian~Matthé, Peter~Neuhaus, Gerhard~Fettweis}
\thanks{Manuscript received February 22, 2018.}
\thanks{W. Dias, D. Gaspar and L. Mendes are with Inatel, Sta. Rita Sapucaí, Brazil. e-mail: {\{wheberth, danilo-gaspar, luciano\}@inatel.br}}
\thanks{M. Chafii, M. Matthé, P. Neuhaus and G. Fettweis are with TU Dresden, Dresden, Germany. e-mail: {\{first name.last name\}}@ifn.et.tu-dresden.de}
\thanks{This work was supported by CNPq-Brasil, Finep/CRR under Grant No. 01.14.0231.00 hosted by Inatel, and by 5G-RANGE Brazil-Europe joint project.}
}

\IEEEoverridecommandlockouts

\begin{document}
\onecolumn 
\Huge{\textbf{IEEE Copyright Notice}}
\Large
\vspace*{3\baselineskip}
\begin{flushleft}

Copyright (c) 2018 IEEE

Personal use of this material is permitted. Permission from IEEE must be obtained for all other uses, in any current or future media, including reprinting/republishing this material for advertising or promotional purposes, creating new collective works, for resale or redistribution to servers or lists, or reuse of any copyrighted component of this work in other works.
\vspace*{3\baselineskip}

	\textbf{Published in:}
	2018 European Conference on Networks and Communications (EuCNC), 18-21 June, 2018, Ljubljana, Slovenia
	\vspace*{1\baselineskip}
	
	\textbf{DOI:} 10.1109/EuCNC.2018.8443268
	\vspace*{1\baselineskip}
		
	\textbf{Print at:}\url{https://ieeexplore.ieee.org/document/8443268}
	\vspace*{1\baselineskip}
	
	\textbf{Cite as:}
	\vspace*{0.5\baselineskip}
		
		\normalsize
		\begin{tabular}{|l|}
			\hline
			 Wheberth Dias, Danilo Garspar, Luciano Mendes, Marwa Chafii, Maximilian Matthé,\\Peter Neuhaus and Gerhard Fettweis, ``Performance Analysis of a 5G Transceiver\\Implementation for Remote Areas Scenarios.'' 2018 European Conference on\\Networks and Communications (EuCNC). IEEE, 2018. \\
			\hline
		\end{tabular}
		\vspace*{2\baselineskip}

		\Large
		\textbf{BibTex:}
		\vspace*{0.5\baselineskip}
		\normalsize
			
\begin{tabular}{|l|}
\hline
@inproceedings\{dias2018performance,\\
	\hskip 5mm title=\{Performance Analysis of a 5G Transceiver Implementation for Remote Areas Scenarios\},\\
	\hskip 5mm author=\{Wheberth Dias, Danilo Gaspar, Luciano Mendes, Marwa Chafii,  Maximilian Matth\{\textbackslash'e\},\\
	\hskip 5mm Peter Neuhaus and Gerhard Fettweis \},\\
	\hskip 5mm booktitle=\{2018 European Conference on Networks and Communications (EuCNC)\},\\
	\hskip 5mm pages=\{363--367\},\\
	\hskip 5mm year=\{2018\},\\
	\hskip 5mm organization=\{IEEE\}\\
\}\\
\hline
\end{tabular}

\end{flushleft}

\normalsize
\clearpage
\twocolumn
	
\maketitle
\input{acronyms.tex}
\input{abstract.tex}
\input{keywords.tex}
\input{1.tex}
\input{2.tex}
\input{3.tex}

\input{4.tex}
\bibliographystyle{IEEEtran}
\bibliography{library}

\end{document}

%% file: preamble.tex
\usepackage[T1]{fontenc}
\usepackage[utf8]{inputenc}
\usepackage{hyperref}
\usepackage{ifthen}
\usepackage{amsmath}
\usepackage{amssymb}
\usepackage{comment}
\usepackage{booktabs}
\usepackage{graphicx}
\usepackage[nolist]{acronym}
\usepackage[normalem]{ulem}
\usepackage[caption=false,font=footnotesize]{subfig}
\usepackage[font=footnotesize]{caption}
\usepackage{lipsum} 

\usepackage{hyphenat}
\usepackage{tikz}
\usepackage{pgfplots}
\pgfplotsset{compat=1.10}
\usepackage{float}

\usetikzlibrary{arrows,calc,positioning}

\usepgfplotslibrary{groupplots}

\iftrue  

\else

\fi

\newcommand{\stkout}[1]{\ifmmode\text{\sout{\ensuremath{#1}}}\else\sout{#1}\fi}

\usepackage{xcolor}

\usepackage[makeroom]{cancel}


%% file: acronyms.tex
\begin{acronym}
  \acro{1G}{First Generation}
  \acro{2G}{Second Generation}
  \acro{3G}{Third Generation}
  \acro{4G}{Fourth Generation}
  \acro{5G}{Fifth Generation}
  \acro{5GRA}{Remote Areas applications}
  \acro{5GPHY}{5G Physical Layer}
  \acro{ADC}{analogic-to-digital converter}
  \acro{AGC}{automatic gain control}
  \acro{ASIP}{Application Specific Integrated Processors}
  \acro{AWGN}{additive white Gaussian noise}
  \acro{BDTM}{burst data transfer mode}
  \acro{BER}{bit error rate}
  \acro{BS}{base station}
  \acro{CDTM}{continuous data transfer mode}
  \acro{CFO}{Carrier Frequency Offset}
  \acro{CHF}{Characteristic Function}  
  \acro{CoMP} {Cooperative Multi-point}
  \acro{CP}{cyclic prefix}
  \acro{CR}{Cognitive Radio}
  \acro{CS}{cyclic suffix}
  \acro{CSI}{channel state information}
  \acro{CSMA}{carrier sense multiple access}
  \acro{DFT}{discrete Fourier transform}
  \acro{DPD}{digital pre-distortion}
  \acro{DZT}{discrete Zak transform}
  \acro{eMBB}{enhanced mobile broadband}
  \acro{EPC}{evolved packet core}
  \acro{FBMC}{Filter-bank multi-carrier}
  \acro{FDE}{frequency-domain equalizer}
  \acro{FDMA}{frequency division multiple access}
  \acro{FD-OQAM-GFDM}{frequency-domain OQAM-GFDM}
  \acro{FEC}{forward error control}
  \acro{FPGA}{Field Programmable Gate Array}
  \acro{FTN}{Faster than Nyquist}
  \acro{FT}{Fourier transform}
  \acro{FSC}{frequency-selective channel}
  \acro{GFDM}{Generalized Frequency Division Multiplexing}
  \acro{GS-GFDM}{guard-symbol GFDM}
  \acro{HPA}{high power amplifier}
  \acro{IBI}{inter-block interference}  
  \acro{ICI}{inter-carrier interference}
  \acro{IDFT}{Inverse Discrete Fourier Transform}
  \acro{IFI}{inter-frame interference}
  \acro{IMS}{IP multimedia subsystem}
  \acro{IoT}{Internet of Things}
  \acro{IP}{Internet Protocol}
  \acro{IQ}{in-phase and quadrature}
  \acro{ISI}{inter-symbol interference}
  \acro{IUI}{inter-user interference}
  \acro{KPI}{key performance indicator}
  \acro{LDPC}{low density check parity code}
  \acro{LLR}{log-likelihood ratio}
  \acro{LMMSE}{linear minimum mean square error}
  \acro{LTE}{Long-Term Evolution}
  \acro{LTE-A}{Long-Term Evolution - Advanced}
  \acro{M2M}{Machine-to-Machine}
  \acro{MA}{multiple access}
  \acro{MAC}{medium access control layer}
  \acro{MF}{Matched filter}
  \acro{MIMO}{multiple-input multiple-output}
  \acro{MMSE}{minimum mean square error}
  \acro{MRC}{maximum ratio combiner}
  \acro{MSE}{mean-squared error}
  \acro{mMTC}{massive machine type communication}
  \acro{MTC}{machine type communication}
  \acro{MU}{multi user}
  \acro{NEF}{noise enhancement factor}
  \acro{NFV}{network functions virtualization}
  \acro{OFDM}{Orthogonal Frequency Division Multiplexing}
  \acro{OOB}{out-of-band}
  \acro{OQAM}{Offset Quadrature Amplitude Modulation}
  \acro{PAPR}{peak to average power ratio}
  \acro{PHY}{physical layer}
  \acro{PRBS}{Pseudo Random Bit Sequence}
  \acro{PSD}{Power Spectrum Density}
  \acro{QAM}{quadrature amplitude modulation}
  \acro{QPSK}{quadrature phase shift keying}
  \acro{QoE}{Quality of Experience}
  \acro{QoS}{Quality of Service}
  \acro{RC}{raised-cosine}
  \acro{ROF}{roll-off factor}
  \acro{RRC}{root raised cosine}
  \acro{SC}{single carrier}
  \acro{SC-FDE}{Single Carrier Frequency Domain Equalization}
  \acro{SC-FDMA}{Single Carrier Frequency Domain Multiple Access}
  \acro{SCD}{Successive Cancellation decoding}
  \acro{SDN}{software-defined network}
  \acro{SDR}{software-defined radio}
  \acro{SDW}{software-defined waveform}
  \acro{SEP}{symbol error probability}
  \acro{SER}{symbol error rate}
  \acro{SIC}{successive interference cancellation}
  \acro{SISO}{single-input single-output}
  \acro{SMS}{Short Message Service}
  \acro{SNR}{signal-to-noise ratio}
  \acro{ST}{space-time}
  \acro{STO}{Symbol Timing Offset}
  \acro{STC}{space time code}
  \acro{STFT}{short-time Fourier transform}
  \acro{TD-OQAM-GFDM}{time-domain OQAM-GFDM}
  \acro{TR-STC}{time-reversal space-time coding}
  \acro{TR-STC-GFDMA}{TR-STC Generalized Frequency Division Multiple Access}
  \acro{TVC}{time-variant channel}
  \acro{TVWS}{TV white space}
  \acro{UHF}{Ultra High Frequency}
  \acro{URLL}{ultra-reliable low latency}
  \acro{V2V}{vehicle-to-vehicle}
  \acro{VHF}{Very High Frequency}
  \acro{V-OFDM}{Vector OFDM}
  \acro{ZF}{zero-forcing}
  \acro{W-GFDM}{windowed GFDM}
  \acro{WHT}{Walsh-Hadamard Transform}
  \acro{WLAN}{wireless Local Area Network}
  \acro{WLE}{widely linear equalizer}
  \acro{WLP}{wide linear processing}
  \acro{WRAN}{Wireless Regional Area Network}
  \acro{WSN}{wireless sensor networks}
\end{acronym}

%% file: abstract.tex
\begin{abstract}
Fifth generation of mobile communication networks will support a large set of new services and applications. One important use case is the remote area coverage for broadband Internet access. This use case has significant social and economical impact, since a considerable percentage of the global population living in low populated area does not have Internet access and the communication infrastructure in rural areas can be used to improve agrobusiness productivity. The aim of this paper is to analyze the performance of a 5G for Remote Areas transceiver, implemented on field programmable gate array based hardware for real-time processing. This transceiver employs the latest digital communication techniques, such as generalized frequency division multiplexing waveform combined with 2 by 2 multiple-input multiple-output diversity scheme and  polar channel coding. The performance of the prototype is evaluated regarding its out-of-band emissions and bit error rate under AWGN channel.

\end{abstract}

%% file: keywords.tex

%% file: 1.tex
\section{Introduction}\label{sec:1}
\IEEEpubidadjcol 
\ac{5G} Networks are being pointed as the next revolution in mobile communications \cite{flexibility_autonomy_5G}, which will support several new services and applications. Several scientific and industrial efforts are currently being made in order to define the new radio interface for \ac{eMBB}, \ac{URLL} and \ac{mMTC} applications \cite{5G_Research}.

However, one important scenario with huge social and economical impact is not being widely discussed by academia and industry: the \ac{5G} operation mode for remote areas. This scenario has very specific requirements and challenges \cite{5G_remote_challenges}. Since the user density in rural areas is small, each \ac{BS} shall cover a large area, leading to long channel delay profiles. \ac{VHF} and \ac{UHF} frequency bands can be exploited due to their good propagation properties. But, since these bands are also used for other services, i.e., digital television, \ac{5G} for remote areas \ac{PHY} must employ a waveform with low \ac{OOB} emission, allowing for \ac{CR} technologies and secondary network methodologies \cite{CR} to be employed. Also, the \ac{PHY} must provide high robustness against the channel impairments, using the state-of-the-art channel coding \cite{polar_codes}.

In \cite{Ngara_paper}, the authors propose a high spectrum efficient \ac{PHY} based on \ac{MU}-\ac{MIMO} for \ac{OFDM} \cite{MIMO-OFDM}. However, mechanisms that allow the coexistence with other legacy networks, reduce the \ac{OOB} emissions and deal with large channel delay profiles are not considered. In \cite{IoT_remote_areas}, the authors present a \ac{MAC}-\ac{PHY} solution for supporting \ac{IoT} and \ac{MTC} in rural areas, but without considering other important applications, such as broadband Internet access and latency sensitive services \cite{5G_low_income}. 

The aim of this paper is to evaluate the performance of a real-time implementation of a transceiver and frame structure, conceived to support 5G services in remote areas. The transceiver is based on a flexible novel modulation scheme, named \ac{GFDM} \cite{gfdm_5th_cel_network}. \ac{GFDM} can be tailored to efficiently use the \ac{CP} and \ac{CS} in severe multipath channels. It also presents very low \ac{OOB} emissions and can be combined with \ac{MIMO} techniques to provide robustness and spectrum efficiency \cite{TRSTC-GFDM} \cite{MIMO-GFDM}. Polar encoder and decoder \cite{polar} \cite{polar_codes} have also been implemented for real-time processing. In this paper, the \ac{BER} performance of \ac{SISO} and \ac{MIMO} \ac{GFDM} system with polar channel coding under \ac{AWGN} channel will be presented. The \ac{OOB} emission will also be analyzed. In order to identify the impact of the hardware implementation into the system performance, two approaches are used for the measurements: i) noiseless channel estimation preambles, where noise is added only to the synchronization preamble and to the GFDM blocks and; ii) noisy channel estimation preambles, where noise is added to the entire GFDM frame. Theoretical and simulation \ac{BER} curves are used as reference.

The remainder of this paper is organized as follows: Section \ref{sec:2} describes the real-time transceiver implementation, while Section \ref{sec:3} brings the systems performance analysis and Section \ref{sec:4} concludes this paper. 

%% file: 2.tex
\section{Transceiver Description}\label{sec:2}
\IEEEpubidadjcol 

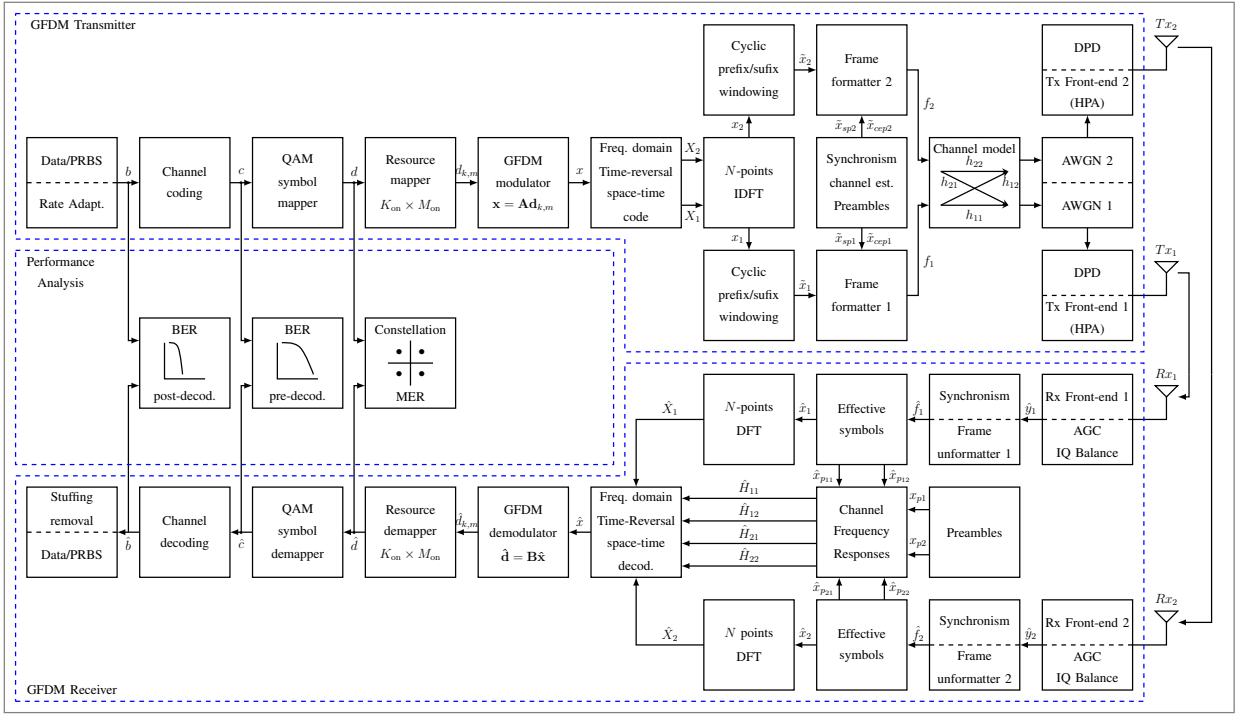
\begin{figure*}[!ht]
	\centering
    \captionsetup{justification=centering}
	\scalebox{0.6}{\input{gfdm_modem_model_block_diagram.tex}}	    
	\caption{Transceiver GFDM block diagram.}	
	\label{fig:transceiver_block_diagram}
\end{figure*}

Figure \ref{fig:transceiver_block_diagram} presents the block diagram of the proposed \ac{5G} transmitter and receiver. 
The \ac{5G} transceiver is able to operate in \ac{CDTM}, where the time-frequency resource is continuously occupied, or in \ac{BDTM}, where the time-frequency resource is only allocated when there are useful data from users to be transmitted. Hereafter, a description of the main transceiver modules is provided.

On the transmitter side, a rate adaptation is performed through insertion of stuffing bits when the transceiver operates in \ac{CDTM} and the income throughput is smaller than its capacity. An internal \ac{PRBS} can be selected in order to allow system performance evaluation.

Next to rate adaptation, income data are encoded by the Channel Coding block, responsible for adding redundancy information to the sequence, seeking to improve the robustness of the system against impairments introduced by the channel. Polar code was chosen to compose the first transceiver version due to its low complexity and superior performance considering short code words \cite{PC5G_contender}. Polar code is implemented under \ac{SCD}, using the shortening technique described in \cite{PC_shortening} with selectable code rates of 1/2, 2/3, 3/4 and 5/6.

The encoded data sequence is mapped according to a \ac{QAM} constellation. The available configurations are 4-QAM, 16-QAM, 64-QAM or 256-QAM.

\ac{GFDM} waveform can be seen as a time-frequency resource grid, arranged in $K$ sub-carriers in frequency and $M$ subsequent sub-symbols in time. Thereby, a total of $N=KM$ data symbols are transmitted in a \ac{GFDM} block, which is given by
\begin{equation}\label{eq:xn}
x[n] = \sum_{k=0}^{K-1}\sum_{m=0}^{M-1}d_{k,m}g\left[\left<n-mK\right>_N\right]{e}^{j2\pi\frac{k}{K}n},
\end{equation}
\noindent
where $d_{k,m}$ is the data symbol carried by the $kth$ sub-carrier at the $mth$ sub-symbol, $g[n]$ is the transmission prototype pulse, $\left<\cdot\right>_N$ represents the modulo $N$ operator and $n=0,1,..,N-1$.

Notably, each data symbol is transmitted in a version of the prototype pulse that is circularly shifted in time and frequency. These pulses can be arranged in a modulation matrix, given by 
\begin{equation}\label{eq:Amat}
	\mathbf{A}=
    \begin{bmatrix}
    	\mathbf{g}_{0,0} & \mathbf{g}_{1,0} \dots \mathbf{g}_{K-1,0} & \mathbf{g}_{0,1} \dots \mathbf{g}_{K-1,M-1}
    \end{bmatrix},
\end{equation}
\noindent
where
\begin{equation}\label{eq:g}
\mathbf{g}_{k,m} = g\left[\left<n-mK\right>_N\right]{e}^{j2\pi\frac{k}{K}n}
\end{equation}
\noindent
is a vector that contains the samples from the circular shifted versions of the prototype pulse. Thus, (\ref{eq:xn}) can be rewritten using the matrix notation as
\begin{equation}\label{eq:x}
\mathbf{x} = \mathbf{Ad},
\end{equation}
\noindent
where $\mathbf{d}=\left(\mathbf{d}^\text{T}_0\cdots\mathbf{d}^\text{T}_{M-1}\right)^\text{T}$ and $\mathbf{d}_m=\left(\mathbf{d}_{0,m}\cdots\mathbf{d}_{K-1,m}\right)^\text{T}$.

The \ac{GFDM} signal can achieve a reduced \ac{OOB} emission by virtue of the circular filtering and the characteristics of the transmission pulse. As the future mobile network will have to coexist with other legacy technologies without introducing interference, the low \ac{OOB} emission is an important feature for \ac{5GPHY}.

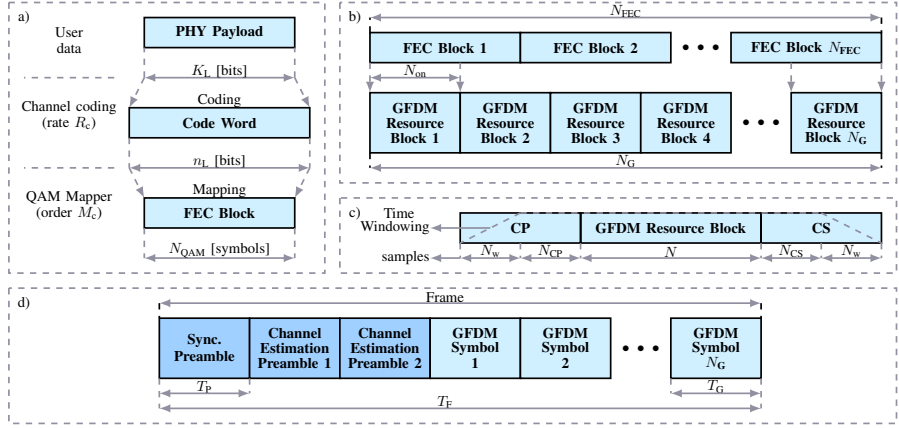
\begin{figure*}[h!tb]
	\centering
    \captionsetup{justification=centering}
	\scalebox{0.8}{\input{frame_structure_diagram.tex}}	    
	\caption{Frame structure diagram \cite{ferreira2017gfdm}.}	
	\label{fig:frame_structure_diagram}
\end{figure*}

A \ac{TR-STC} scheme is applied to the modulated signal, resulting in two correlated versions of the transmission signal irradiated towards the two receiver antennas installed several wave lengths away. This technique permits the receiver to explore the diversity gain from the multi-path channel, heightening the overall system performance. A \ac{CP} and a \ac{CS} are added to the transmission signals aiming to protect the symbols from the \ac{ISI} introduced by dispersive channels. A time window can also be applied to symbols in order to smooth the abrupt transitions between \ac{GFDM} blocks and improve the reduction of the \ac{OOB} emissions.

The frame structure proposed for this transceiver is configurable and allows different operation modes in order to cover the requirements of the \ac{5G} scenarios. The frame formatter block multiplexes the user data with the synchronization and channel estimation preambles, used by the receiver to recover synchronism and perform the channel equalization. Figure \ref{fig:frame_structure_diagram} \cite{ferreira2017gfdm} details the frame structure.
Most of transceiver parameters, such as coding rate, modulation order, number of active sub-carriers/sub-symbols and occupied bandwidth, \ac{CP}/\ac{CS}/Windowing length and the rate of synchronization/channel estimation preambles are allowed to be configured according with channel conditions and application requirements. 


The transceiver implementation also comes with a \ac{DPD} in order to reduce the effect of the non-linear distortions introduced by the \ac{HPA} and consequently spectral regrowth, caused by the inherent high \ac{PAPR} of multi-carrier systems, once the high amplitude peaks can lead the \ac{HPA} to its non-linear region, resulting in high \ac{OOB} emissions and \ac{ICI}.

At the receiver side, right after the RF front-end, the \ac{AGC} block operates to normalize the input level and to properly exploit the dynamic range of the \ac{ADC}. The IQ balance block removes the \ac{IQ} imbalance introduced by the RF chain. Time and frequency synchronization is performed using the transmitted preamble and its locally stored replica, where frame structure timing is recovered and its portions are identified. Following synchronization, the channel frequency responses based on the received preambles are estimated. The channel frequency responses are then used by the \ac{ST} decoder to combine the received \ac{GFDM} blocks from the multiple antennas, achieving a diversity gain. Since the transceiver employs two transmit and receive antennas, the system can achieve a diversity of order 4 (\ac{MIMO} 2x2).

The \ac{ST} decoder equalizes the combined signal and delivers it to the \ac{GFDM} demodulator to recover the transmitted \ac{QAM} sequence. The current version of the transceiver employs a zero-forcing demodulator, which uses an inverse of the modulation matrix as follows
\begin{equation}\label{eq:dhat}
	\mathbf{\hat{d}}=\mathbf{B}\mathbf{y}_{\text{eq}},
\end{equation}
\noindent
where $\mathbf{y}_{\text{eq}}$ is the equalized received signal at the output of the	\ac{ST} decoder and
\begin{equation}\label{eq:Bmat}
	\mathbf{B}=\mathbf{A}^{-1}
\end{equation}
\noindent
is the demodulation matrix.

The recovered \ac{QAM} data feeds the symbol de-mapper block and the resulting bit sequence is used by the channel decoder block for correcting errors introduced by the channel. Then bit stuffing removal is applied in order to recover only the relevant transmitted information.

%% file: gfdm_modem_model_block_diagram.tex
\begin{tikzpicture}[thick,scale=1, every node/.style={scale=0.7}]
\large
\draw  (-5.25,3.75) rectangle (-3.25,1.75);
\node at (-4.25,3.25) {Data/PRBS};
\draw[dashed] (-5.25,2.75) -- (-3.25,2.75);
\node at (-4.25,2.25) {Rate Adapt.};

\draw  (-2.75,3.75) rectangle (-0.75,1.75);
\node at (-1.75,3) {Channel};
\node at (-1.75,2.5) {coding};

\draw  (-0.25,3.75) rectangle (1.75,1.75);
\node at (0.75,3.25) {QAM};
\node at (0.75,2.75) {symbol};
\node at (0.75,2.25) {mapper};

\draw  (2.25,3.75) rectangle (4.25,1.75);
\node at (5.75,3.25) {GFDM};
\node at (5.75,2.75) {modulator};
\node at (5.75,2.25) {$\mathbf{x}=\mathbf{A}\mathbf{d}_{k,m}$};

\draw  (4.75,3.75) rectangle (6.75,1.75);
\node at (-4,6.25) {GFDM Transmitter};

\node at (3.25,3.25) {Resource};
\node at (3.25,2.75) {mapper};
\node at (3.25,2.25) {$K_\mathrm{on} \times M_\mathrm{on}$};

\draw  (7.25,3.75) rectangle (9.25,1.75);
\node at (8.25,3.5) {Freq. domain};
\node at (8.25,3) {Time-reversal};
\node at (8.25,2.5) {space-time};
\node at (8.25,2) {code};

\draw  (12.25,6.25) rectangle (14.25,4.25);
\node at (13.25,5.5) {Frame};
\node at (13.25,5) {formatter 2};

\draw  (9.75,6.25) rectangle (11.75,4.25);
\node at (10.75,5.75) {Cyclic};
\node at (10.75,5.25) {prefix/sufix};
\node at (10.75,4.75) {windowing};

\draw  (9.75,1.25) rectangle (11.75,-0.75);
\node at (10.75,0.75) {Cyclic};
\node at (10.75,0.25) {prefix/sufix};
\node at (10.75,-0.25) {windowing};

\draw  (17.25,6.25) rectangle (19.25,4.25);
\node at (18.25,5.75) {DPD};
\draw[dashed] (17.25,5.25) -- (19.25,5.25);
\node at (18.25,5) {Tx Front-end 2};
\node at (18.25,4.5) {(HPA)};
\draw (19.25,5.25) -- (20,5.25) -- (20,5.75) -- (19.75,6) -- (20.25,6) -- (20,5.75);

\draw  (17.25,1.25) rectangle (19.25,-0.75);
\node at (18.25,0.75) {DPD};
\draw[dashed] (17.25,0.25) -- (19.25,0.25);
\node at (18.25,0) {Tx Front-end 1};
\node at (18.25,-0.5) {(HPA)};
\draw (19.25,0.25) -- (20,0.25) -- (20,0.75) -- (19.75,1) -- (20.25,1) -- (20,0.75);

\draw  (14.75,3.75) rectangle (16.75,1.75);
\draw [-stealth](15,2.25) -- (16.5,2.25);
\draw [-stealth](15,2.25) -- (16.5,3);
\draw [-stealth](15,3) -- (16.5,3);
\draw [-stealth](15,3) -- (16.5,2.25);
\node [rotate=0,below] at (15.75,2.25) {$h_{11}$};
\node [rotate=0,right] at (16.25,2.75) {$h_{12}$};
\node [rotate=0,left] at (15.5,2.75) {$h_{21}$};
\node [rotate=0,above] at (15.75,3) {$h_{22}$};
\node at (15.75,3.5) {Channel model};

\draw  (17.25,-1.5) rectangle (19.25,-3.5);
\node at (18.25,-2) {Rx Front-end 1};
\draw[dashed] (17.25,-2.5) -- (19.25,-2.5);
\node at (18.25,-2.75) {AGC};
\node at (18.25,-3.25) {IQ Balance};
\draw (19.25,-2.5) -- (20,-2.5) -- (20,-2) -- (19.75,-1.75) -- (20.25,-1.75) -- (20,-2);

\draw  (17.25,-6.5) rectangle (19.25,-8.5);
\node at (18.25,-7) {Rx Front-end 2};
\draw[dashed] (17.25,-7.5) -- (19.25,-7.5);
\node at (18.25,-7.75) {AGC};
\node at (18.25,-8.25) {IQ Balance};
\draw (19.25,-7.5) -- (20,-7.5) -- (20,-7) -- (19.75,-6.75) -- (20.25,-6.75) -- (20,-7);

\draw  (14.75,-1.5) rectangle (16.75,-3.5);
\node at (15.75,-2) {Synchronism};

\draw  (14.75,-6.5) rectangle (16.75,-8.5);
\node at (15.75,-7) {Synchronism};

\draw  (12.25,-1.5) rectangle (14.25,-3.5);
\node at (15.75,-2.75) {Frame};
\node at (15.75,-3.25) {unformatter 1};

\draw  (12.25,-4) rectangle (14.25,-6);
\node at (13.25,-4.5) {Channel};
\node at (13.25,-5) {Frequency};
\node at (13.25,-5.5) {Responses};

\draw  (12.25,-6.5) rectangle (14.25,-8.5);
\node at (15.75,-7.75) {Frame};
\node at (15.75,-8.25) {unformatter 2};

\draw[dashed] (14.75,-2.5) -- (16.75,-2.5);
\draw[dashed] (14.75,-7.5) -- (16.75,-7.5);

\draw  (7.25,-4) rectangle (9.25,-6);
\node at (8.25,-4.25) {Freq. domain};
\node at (8.25,-4.75) {Time-Reversal};
\node at (8.25,-5.25) {space-time};
\node at (8.25,-5.75) {decod.};

\draw  (4.75,-4) rectangle (6.75,-6);
\node at (5.75,-4.5) {GFDM};
\node at (5.75,-5) {demodulator};
\node at (5.75,-5.5) {$\mathbf{\hat{d}}=\mathbf{B}\mathbf{\hat{x}}$};

\draw  (2.25,-4) rectangle (4.25,-6);
\node at (3.25,-4.5) {Resource};
\node at (3.25,-5) {demapper};
\node at (3.25,-5.5) {$K_\mathrm{on} \times M_\mathrm{on}$};

\draw  (-0.25,-4) rectangle (1.75,-6);
\node at (0.75,-4.5) {QAM};
\node at (0.75,-5) {symbol};
\node at (0.75,-5.5) {demapper};

\draw  (-2.75,-4) rectangle (-0.75,-6);
\node at (-1.75,-4.75) {Channel};
\node at (-1.75,-5.25) {decoding};

\draw  (-2.75,-0.25) rectangle (-0.75,-2.25);
\node at (-1.75,-0.5) {BER};

\node at (-1.75,-2) {post-decod.};

\draw  (-0.25,-0.25) rectangle (1.75,-2.25);
\node at (0.75,-0.5) {BER};

\node at (0.75,-2) {pre-decod.};

\draw  (2.25,-0.25) rectangle (4.25,-2.25);
\node at (3.25,-0.5) {Constellation};
\draw (3.25,-0.75) -- (3.25,-1.75);
\draw (2.75,-1.25) -- (3.75,-1.25);
\node at (3,-1) {\textbullet};
\node at (3,-1.5) {\textbullet};
\node at (3.5,-1) {\textbullet};
\node at (3.5,-1.5) {\textbullet};
\node at (3.25,-2) {MER};

\draw[dashed, blue] (-5.5,6.5) -- (19.5,6.5) -- (19.5,-1) -- (8,-1) -- (8,1.5) -- (-5.5,1.5) -- (-5.5,6.5);
\draw [-latex](-3.25,2.75) -- (-2.75,2.75);
\draw [-latex](-0.75,2.75) -- (-0.25,2.75);
\draw [-latex](1.75,2.75) -- (2.25,2.75);
\draw [-latex](4.25,2.75) -- (4.75,2.75);
\draw [-latex](6.75,2.75) -- (7.25,2.75);

\draw [-latex](11.75,5.25) -- (12.25,5.25);
\draw [-latex](11.75,0.25) -- (12.25,0.25);
\draw [-](20.25,0.75) -- (20.5,0.75) -- (20.5,-1.5);
\draw [-](20.25,5.75) -- (21,5.75) -- (21,-1.5);
\draw [-latex](20.5,-1.5) -- (20.5,-2) -- (20.25,-2);
\draw [-latex](21,-1.5) -- (21,-7) -- (20.25,-7);
\draw [-latex](17.25,-2.5) -- (16.75,-2.5);
\draw [-latex](17.25,-7.5) -- (16.75,-7.5);
\draw [-latex](14.75,-2.5) -- (14.25,-2.5);
\draw [-latex](14.75,-7.5) -- (14.25,-7.5);

\draw [-latex](7.25,-5) -- (6.75,-5);
\draw [-latex](4.75,-5) -- (4.25,-5);
\draw [-latex](2.25,-5) -- (1.75,-5);
\draw [-latex](-0.25,-5) -- (-0.75,-5);
\draw [-latex](-3,2.75) -- (-3,-0.75) -- (-2.75,-0.75);
\draw [-latex](-3,-5) -- (-3,-1.75) -- (-2.75,-1.75);
\draw [-latex](-0.5,2.75) -- (-0.5,-0.75) -- (-0.25,-0.75);
\draw [-latex](-0.5,-5) -- (-0.5,-1.75) -- (-0.25,-1.75);
\draw [-latex](2,2.75) -- (2,-0.75) -- (2.25,-0.75);
\draw [-latex](2,-5) -- (2,-1.75) -- (2.25,-1.75);
\node at (-3,2.75) {\tiny{\textbullet}};
\node at (-0.5,2.75) {\tiny{\textbullet}};
\node at (2,2.75) {\tiny{\textbullet}};
\node at (-0.5,-5) {\tiny{\textbullet}};
\node at (2,-5) {\tiny{\textbullet}};

\draw  (9.75,3.75) rectangle (11.75,1.75);
\node at (10.75,3) {$N$-points};
\node at (10.75,2.5) {IDFT};
\node at (2,3) {$d$};
\node at (4.5,3) {$d_{k,m}$};
\node at (7,3) {$x$};

\draw[-latex] (9.25,3.25) -- (9.75,3.25);
\draw[-latex] (9.25,2.25) -- (9.75,2.25);
\node at (9.5,3.5) {$X_2$};
\node at (9.5,2) {$X_1$};

\node[left] at (10.75,4) {$x_2$};
\node[left] at (10.75,1.5) {$x_1$};
\node[right] at (14.5,4.5) {$f_2$};
\node[right] at (14.5,1) {$f_1$};
\node[above] at (12,5.25) {$\tilde{x}_2$};
\node[above] at (12,0.25) {$\tilde{x}_1$};
\node at (20,6.25) {$Tx_2$};
\node at (20,1.25) {$Tx_1$};
\node at (-3,3) {$b$};
\node at (-0.5,3) {$c$};
\node at (13.25,-2.25) {Effective};
\node at (13.25,-2.75) {symbols};
\node at (13.25,-7.25) {Effective};
\node at (13.25,-7.75) {symbols};
\draw  (14.75,-4) rectangle (16.75,-6);
\node at (15.75,-5) {Preambles};
\draw  (12.25,1.25) rectangle (14.25,-0.75);
\node at (13.25,0.5) {Frame};
\node at (13.25,0) {formatter 1};
\draw[-latex] (14.25,5.25) -- (14.5,5.25) -- (14.5,3.25) -- (14.75,3.25);
\draw[-latex] (14.25,0.25) -- (14.5,0.25) -- (14.5,2.25) -- (14.75,2.25);
\draw[-latex] (10.75,3.75) -- (10.75,4.25);
\draw[-latex] (10.75,1.75) -- (10.75,1.25);
\draw  (12.25,3.75) rectangle (14.25,1.75);
\node at (13.25,3.25) {Synchronism};
\node at (13.25,2.75) {channel est.};
\node at (13.25,2.25) {Preambles};
\draw[-latex] (13.25,3.75) -- (13.25,4.25);
\draw[-latex] (13.25,1.75) -- (13.25,1.25);
\node[left] at (13.25,4) {$\tilde{x}_{sp2}$};
\node[right] at (13.25,4) {$\tilde{x}_{cep2}$};
\node[left] at (13.25,1.5) {$\tilde{x}_{sp1}$};
\node[right] at (13.25,1.5) {$\tilde{x}_{cep1}$};
\draw[-latex] (14.75,-4.5) -- (14.25,-4.5);
\node at (14.5,-4.25) {$x_{p1}$};
\draw[-latex] (14.75,-5.5) -- (14.25,-5.5);
\node at (14.5,-5.25) {$x_{p2}$};
\draw[-latex] (12.75,-3.5) -- (12.75,-4);
\draw[-latex] (13.75,-3.5) -- (13.75,-4);
\draw[-latex] (12.75,-6.5) -- (12.75,-6);
\draw[-latex] (13.75,-6.5) -- (13.75,-6);
\node[left] at (12.75,-3.75) {$\hat{x}_{p_{11}}$};
\node[right] at (13.75,-3.75) {$\hat{x}_{p_{12}}$};
\node[left] at (12.75,-6.25) {$\hat{x}_{p_{21}}$};
\node[right] at (13.75,-6.25) {$\hat{x}_{p_{22}}$};
\draw  (9.75,-1.5) rectangle (11.75,-3.5);
\node at (10.75,-2.75) {DFT};
\node at (10.75,-2.25) {$N$-points};


\draw  (9.75,-6.5) rectangle (11.75,-8.5);
\node at (10.75,-7.75) {DFT};
\node at (10.75,-7.25) {$N$ points};
\draw[-latex] (9.75,-2.5) -- (8.25,-2.5) -- (8.25,-4);
\draw[-latex] (9.75,-7.5) -- (8.25,-7.5) -- (8.25,-6);
\draw[-latex] (12.25,-4.25) -- (9.25,-4.25);
\draw[-latex] (12.25,-4.75) -- (9.25,-4.75);
\draw[-latex] (12.25,-5.25) -- (9.25,-5.25);
\draw[-latex] (12.25,-5.75) -- (9.25,-5.75);
\node at (10.75,-4) {$\hat{H}_{11}$};
\node at (10.75,-4.5) {$\hat{H}_{12}$};
\node at (10.75,-5) {$\hat{H}_{21}$};
\node at (10.75,-5.5) {$\hat{H}_{22}$};
\node at (9,-2.25) {$\hat{X}_1$};
\node at (9,-7.25) {$\hat{X}_2$};
\draw[-latex] (12.25,-2.5) -- (11.75,-2.5);

\draw[-latex] (12.25,-7.5) -- (11.75,-7.5);
\node at (12,-2.25) {$\hat{x}_1$};
\node at (12,-7.25) {$\hat{x}_2$};
\node at (7,-4.75) {$\hat{x}$};
\node at (4.5,-4.75) {$\hat{d}_{k,m}$};
\node at (2,-5.25) {$\hat{d}$};
\node at (-0.5,-5.25) {$\hat{c}$};
\node[below] at (-3,-5) {$\hat{b}$};
\node at (20,-1.5) {$Rx_1$};
\node at (20,-6.5) {$Rx_2$};
\node at (17,-2.25) {$\hat{y}_1$};
\node at (17,-7.25) {$\hat{y}_2$};
\node at (14.5,-2.25) {$\hat{f}_1$};
\node at (14.5,-7.25) {$\hat{f}_2$};
\draw[dashed,blue] (-5.5,-8.75) -- (19.5,-8.75) -- (19.5,-1.25) -- (8,-1.25) -- (8,-3.75) -- (-5.5,-3.75) -- (-5.5,-8.75);
\node at (-4.25,-8.5) {GFDM Receiver};
\draw (-2.2,-0.8) -- (-2.2,-1.6) -- (-1.3,-1.6);
\draw  plot[smooth, tension=.7] coordinates {(-2.1,-0.8)  (-1.9,-0.9) (-1.8,-1.5)};
\draw (0.3,-0.8) -- (0.3,-1.6) -- (1.2,-1.6);
\draw  plot[smooth, tension=.7] coordinates {(0.4,-0.8) (0.8,-0.9) (1.1,-1.5)};
\draw[gray]  (-5.75,6.75) rectangle (21.5,-9);

\draw  (-5.25,-4) rectangle (-3.25,-6);
\draw[dashed] (-5.25,-5) -- (-3.25,-5);
\node at (-4.25,-4.25) {Stuffing};
\node at (-4.25,-4.75) {removal};
\node at (-4.25,-5.5) {Data/PRBS};
\draw[-latex] (-2.75,-5) -- (-3.25,-5);

\draw[-latex] (16.75,3.25) -- (17.25,3.25);
\draw[-latex] (16.75,2.25) -- (17.25,2.25);
\draw  (17.25,3.75) rectangle (19.25,1.75);
\draw[dashed] (17.25,2.75) -- (19.25,2.75);
\node at (18.25,3.25) {AWGN 2};
\node at (18.25,2.25) {AWGN 1};
\draw[-latex]  (18.25,3.75) -- (18.25,4.25);
\draw[-latex]  (18.25,1.75) -- (18.25,1.25);

\draw[dashed,blue] (-5.5,1.25) -- (7.75,1.25) -- (7.75,-3.5) -- (-5.5,-3.5) -- (-5.5,1.25);
\node at (-4.5,1) {Performance};
\node at (-4.5,0.5) {Analysis};
\end{tikzpicture}

%% file: frame_structure_diagram.tex
\definecolor{fill1}{HTML}{D0F0FF}
\definecolor{fill2}{HTML}{A0D0FF}
\definecolor{trace1}{HTML}{9090A0}
\definecolor{txt1}{HTML}{000000}
\definecolor{txt2}{HTML}{000000}

\begin{tikzpicture}[thick,scale=1, every node/.style={scale=0.7},inner sep = 0]
\filldraw[draw=black,fill=fill1]  (-5.5,5.75) rectangle (-3,5.25);
\node [color=txt1] at (-4.25,5.5) {\bf{PHY Payload}};
\draw[color=trace1,latex-latex] (-3,4.75)--(-5.5,4.75);
\node[above,color=txt2] at (-4.25,4.75) {$K_\text{L}$ [bits]};
\draw [dashed,-,color=trace1](-6,4.75) -- (-7.5,4.75);
\node[color=txt2] at (-6.75,5.5) {User};
\node[color=txt2] at (-6.75,5.25) {data};
\draw[color=trace1,dashed,-] (-5.5,4.75)--(-5.5,5.25);
\draw[color=trace1,dashed,-] (-3,4.75)--(-3,5.25);
\draw [color=trace1,dashed,-latex](-5.5,4.75) -- (-5.75,4.25);
\draw [color=trace1,dashed,-latex](-3,4.75) -- (-2.75,4.25);
\node[above,color=txt2] at (-4.25,4.25) {Coding};
\filldraw[draw=black,fill=fill1] (-5.75,4.25) rectangle (-2.75,3.75);
\node [color=txt1] at (-4.25,4) {\bf{Code Word}};
\draw [color=trace1,latex-latex] (-5.75,3.25) -- (-2.75,3.25);
\node[above,color=txt2] at (-4.25,3.25) {$n_\text{L}$ [bits]};
\draw [dashed,-,color=trace1](-6,3.25) -- (-7.5,3.25);
\node[color=txt2] at (-6.75,4.25) {Channel coding};
\node[color=txt2] at (-6.75,4) {(rate $R_\text{c}$)};
\draw [color=trace1,dashed,-] (-5.75,3.75) -- (-5.75,3.25);
\draw [color=trace1,dashed,-] (-2.75,3.75) -- (-2.75,3.25);
\node[above,color=txt2] at (-4.25,2.75) {Mapping};
\draw  [color=trace1,dashed,-latex](-5.75,3.25) -- (-5.5,2.75);
\draw  [color=trace1,dashed,-latex](-2.75,3.25) -- (-3,2.75);
\filldraw[draw=black,fill=fill1] (-5.5,2.75) rectangle (-3,2.25);
\node [color=txt1] at (-4.25,2.5) {\bf{FEC Block}};
\draw [color=trace1,dashed,-] (-5.5,2.25) -- (-5.5,1.75);
\draw [color=trace1,dashed,-] (-3,2.25) -- (-3,1.75);
\draw[color=trace1,latex-latex] (-3,1.75)--(-5.5,1.75);
\node[above,color=txt2] at (-4.25,1.75) {$N_\text{QAM}$ [symbols]};
\node[color=txt2] at (-6.75,2.75) {QAM Mapper};
\node[color=txt2] at (-6.75,2.5) {(order $M_\text{c}$)};
\draw [dashed,color=trace1] (-7.75,6) rectangle (-2.5,1.5);
\node[color=txt2] at (-7.5,5.75) {a)};

\draw [dashed,-](-1.75,5.5) -- (-1.75,5.75);
\draw [dashed,-](6.75,5.75) -- (6.75,5.5);
\draw [color=trace1,latex-latex](-1.75,5.75) -- (6.75,5.75);
\node[above,color=txt2] at (2.5,5.75) {$N_\text{FEC}$};
\filldraw[draw=black,fill=fill1] (-1.75,5.5) rectangle (0.75,5);
\node [color=txt1] at (-0.5,5.25) {\bf{FEC Block 1}};
\filldraw[draw=black,fill=fill1] (0.75,5.5) rectangle (3.25,5);
\node [color=txt1] at (2,5.25) {\bf{FEC Block 2}};
\node at (4.5,4) {\textbullet};
\node at (4.75,4) {\textbullet};
\node at (5,4) {\textbullet};
\filldraw[draw=black,fill=fill1] (4.25,5.5) rectangle (6.75,5);
\node [color=txt1] at (5.5,5.25) {\bf{FEC Block $N_\text{FEC}$}};
\filldraw[draw=black,fill=fill1] (-1.75,4.5) rectangle (-0.25,3.5);
\node [color=txt1] at (-1,4.25) {\bf{GFDM}};
\node [color=txt1] at (-1,4) {\bf{Resource}};
\node [color=txt1] at (-1,3.75) {\bf{Block 1}};
\filldraw[draw=black,fill=fill1] (-0.25,4.5) rectangle (1.25,3.5);
\node [color=txt1] at (0.5,4.25) {\bf{GFDM}};
\node [color=txt1] at (0.5,4) {\bf{Resource}};
\node [color=txt1] at (0.5,3.75) {\bf{Block 2}};
\filldraw[draw=black,fill=fill1] (1.25,4.5) rectangle (2.75,3.5);
\node [color=txt1] at (2,4.25) {\bf{GFDM}};
\node [color=txt1] at (2,4) {\bf{Resource}};
\node [color=txt1] at (2,3.75) {\bf{Block 3}};
\filldraw[draw=black,fill=fill1] (2.75,4.5) rectangle (4.25,3.5);
\node [color=txt1] at (3.5,4.25) {\bf{GFDM}};
\node [color=txt1] at (3.5,4) {\bf{Resource}};
\node [color=txt1] at (3.5,3.75) {\bf{Block 4}};
\node at (3.5,5.25) {\textbullet};
\node at (3.75,5.25) {\textbullet};
\node at (4,5.25) {\textbullet};
\filldraw[draw=black,fill=fill1] (5.25,4.5) rectangle (6.75,3.5);
\node [color=txt1] at (6,4.25) {\bf{GFDM}};
\node [color=txt1] at (6,4) {\bf{Resource}};
\node [color=txt1] at (6,3.75) {\bf{Block $N_\text{G}$}};
\draw [color=trace1,dashed,-latex](-1.75,5) -- (-1.75,4.5);
\draw [color=trace1,dashed,-latex](-0.25,5) -- (-0.25,4.5);
\draw [color=trace1,dashed,-latex](5.25,5) -- (5.25,4.5);
\draw [color=trace1,dashed,-latex](6.75,5) -- (6.75,4.5);
\draw [color=trace1,latex-latex](-1.75,4.75) -- (-0.25,4.75);
\node[above,color=txt2] at (-1,4.75) {$N_\text{on}$};
\draw [dashed,-](-1.75,3.5) -- (-1.75,3.25);
\draw [dashed,-](6.75,3.5) -- (6.75,3.25);
\draw [color=trace1,latex-latex](-1.75,3.25) -- (6.75,3.25);
\node[above,color=txt2] at (2.5,3.25) {$N_\text{G}$};
\draw [dashed,color=trace1] (-2.25,6) rectangle (7,3);
\node[color=txt2] at (-2,5.75) {b)};

\filldraw[draw=black,fill=fill1] (1.75,2.5) rectangle (4.75,2);
\node [color=txt1] at (3.25,2.25) {\bf{GFDM Resource Block}};
\filldraw[draw=black,fill=fill1] (-0.25,2.5) rectangle (1.75,2);
\filldraw[draw=black,fill=fill1] (4.75,2.5) rectangle (6.75,2);
\draw (1.75,2.5) -- (1.75,2);
\draw (4.75,2.5) -- (4.75,2);
\draw [color=trace1,dashed] (0.75,2.5) -- (-0.25,2);
\draw [color=trace1,dashed](0.75,2.5) -- (5.75,2.5);
\draw [color=trace1,dashed] (5.75,2.5) -- (6.75,2);
\node [color=txt1] at (0.75,2.25) {\bf{CP}};
\node [color=txt1] at (5.75,2.25) {\bf{CS}};
\draw [color=trace1,dashed] (-0.25,2) -- (-0.25,1.75);
\draw [color=trace1,dashed] (0.75,2) -- (0.75,1.75);
\draw [color=trace1,dashed] (1.75,2) -- (1.75,1.75);
\draw [color=trace1,dashed] (4.75,2) -- (4.75,1.75);
\draw [color=trace1,dashed] (5.75,2) -- (5.75,1.75);
\draw [color=trace1,dashed] (6.75,2) -- (6.75,1.75);
\draw  [color=trace1,latex-latex](0.75,1.75) -- (-0.25,1.75);
\draw  [color=trace1,latex-latex](1.75,1.75) -- (0.75,1.75);
\draw  [color=trace1,latex-latex](4.75,1.75) -- (1.75,1.75);
\draw  [color=trace1,latex-latex](6.75,1.75) -- (5.75,1.75);
\draw  [color=trace1,latex-latex](5.75,1.75) -- (4.75,1.75);
\node [above,color=txt2] at (0.25,1.75) {$N_\text{w}$};
\node [above,color=txt2] at (1.25,1.75) {$N_\text{CP}$};
\node [above,color=txt2] at (3.25,1.75) {$N$};
\node [above,color=txt2] at (5.25,1.75) {$N_\text{CS}$};
\node [above,color=txt2] at (6.25,1.75) {$N_\text{w}$};
\node [left, color=txt2] at (-1,2.5) {Time};
\node [left, color=txt2] at (-0.75,2.25) {Windowing};
\draw [color=trace1,-latex](0.25,2.25) -- (-0.75,2.25);
\node [left, color=txt2] at (-0.75,1.75) {samples};
\draw [color=trace1,-latex](-0.25,1.75) -- (-0.75,1.75);
\draw [dashed,color=trace1] (-2.25,2.75) rectangle (7,1.5);
\node at (-2,2.5) {c)};

\node [above,color=txt2] at (-0.5,1) {Frame};
\draw  [color=trace1,latex-latex] (4.75,1) -- (-5.25,1);
\draw [dashed](-5.25,1) -- (-5.25,0.75);
\draw [dashed](4.75,1) -- (4.75,0.75);
\filldraw[draw=black,fill=fill2] (-5.25,0.75) rectangle (-3.75,-0.25);
\node [color=txt1] at (-4.5,0.375) {\bf{Sync.}};
\node [color=txt1] at (-4.5,0.125) {\bf{Preamble}};
\filldraw[draw=black,fill=fill2] (-3.75,0.75) rectangle (-2.25,-0.25);
\node [color=txt1] at (-3,0.5) {\bf{Channel}};
\node [color=txt1] at (-3,0.25) {\bf{Estimation}};
\node [color=txt1] at (-3,0) {\bf{Preamble 1}};
\filldraw[draw=black,fill=fill2] (-2.25,0.75) rectangle (-0.75,-0.25);
\node [color=txt1] at (-1.5,0.5) {\bf{Channel}};
\node [color=txt1] at (-1.5,0.25) {\bf{Estimation}};
\node [color=txt1] at (-1.5,0) {\bf{Preamble 2}};
\node at (2.5,0.25) {\textbullet};
\node at (2.75,0.25) {\textbullet};
\node at (3,0.25) {\textbullet};
\filldraw[draw=black,fill=fill1] (-0.75,0.75) rectangle (0.75,-0.25);
\node [color=txt1] at (0,0.5) {\bf{GFDM}};
\node [color=txt1] at (0,0.25) {\bf{Symbol}};
\node [color=txt1] at (0,0) {\bf{1}};
\filldraw[draw=black,fill=fill1] (0.75,0.75) rectangle (2.25,-0.25);
\node [color=txt1] at (1.5,0.5) {\bf{GFDM}};
\node [color=txt1] at (1.5,0.25) {\bf{Symbol}};
\node [color=txt1] at (1.5,0) {\bf{2}};
\filldraw[draw=black,fill=fill1] (3.25,0.75) rectangle (4.75,-0.25);
\node [color=txt1] at (4,0.5) {\bf{GFDM}};
\node [color=txt1] at (4,0.25) {\bf{Symbol}};
\node [color=txt1] at (4,0) {\bf{$N_\text{G}$}};
\draw [color=trace1,dashed](-5.25,-0.25) -- (-5.25,-0.75);
\draw [color=trace1,dashed](-3.75,-0.25) -- (-3.75,-0.5);
\node [above,color=txt2] at (-4.5,-0.5) {$T_\text{P}$};
\draw  [color=trace1,latex-latex](-3.75,-0.5) -- (-5.25,-0.5);
\draw [color=trace1,dashed](3.25,-0.25) -- (3.25,-0.5);
\draw [color=trace1,dashed](4.75,-0.25) -- (4.75,-0.75);
\node [above,color=txt2] at (4,-0.5) {$T_\text{G}$};
\draw  [color=trace1,latex-latex](4.75,-0.5) -- (3.25,-0.5);
\node [above,color=txt2] at (-0.5,-0.75) {$T_\text{F}$};
\draw  [color=trace1,latex-latex] (4.75,-0.75) -- (-5.25,-0.75);
\draw [dashed,color=trace1] (-7.75,1.25) rectangle (7,-1);
\node[color=txt2] at (-7.5,1) {d)};

\end{tikzpicture}

%% file: 3.tex
\section{Performance Analysis}\label{sec:3}

The performance analysis presented in this paper is divided in two \acp{KPI}. The first one is the system \ac{BER}, while the second one is the \ac{OOB} emissions.

\subsection{\ac{BER} performance analysis}

A Monte Carlo process was conducted in order to obtain the overall system performance. The transmission channel is performed by a built-in channel simulator in the transceiver. As a reference for the acquired results, an approximation curve for the theoretical \ac{GFDM} \ac{BER} with \ac{ZF} receiver under \ac{AWGN} is employed and given by \cite{gfdm_5th_cel_network}
\begin{equation}\label{eq:Pb}
	P_\text{b}\simeq\left[\frac{2\left(L-1\right)}{\mu L}\text{erfc}\left(\Gamma\right)-\left(\frac{L-1}{\sqrt{\mu}L}\right)^2\text{erfc}^2\left(\Gamma\right)\right],
\end{equation}
\noindent
where $\text{erfc}(\cdot)$ is the complementary error function, $L=\sqrt{M_\text{c}}$, with $M_\text{c}$ being the \ac{QAM} constellation size, $\mu=\log_2(M_\text{c})$ and
\begin{equation}\label{eq:Gamma}
\Gamma=\frac{3\eta\text{SNR}}{2(L^2-1)\xi}
\end{equation}
\noindent
with $\eta$ representing the frame structure efficiency, considering the \ac{CP}, \ac{CS} and inserted preambles. The \ac{SNR} is in linear scale and $\xi=\sum_{n=0}^{N-1}|\gamma[n]|^2$
is the \ac{NEF} due to the zero-forcing demodulation employed, with $\gamma[n]$ being the samples of the prototype receiver pulse.

The waveform parameters used in the simulations and measurements are described in Table \ref{tab:gfdm_param}. The demodulator used in the simulation is the \ac{ZF}, the same implemented on the hardware prototype.
\begin{table}[!ht]
\centering
\caption{Waveform parameters used in simulation and measurements.}
\label{tab:gfdm_param}
\begin{tabular}{@{}ll@{}}
\toprule
\multicolumn{1}{c}{\textbf{Parameter}} & \multicolumn{1}{c}{\textbf{Value}} \\ \midrule
Constellation size                     & 64-QAM                             \\
$K$                                    & 512                                \\
$M$                                    & 3                                  \\
Prototype pulse                        & \ac{RC} \cite{gfdm_5th_cel_network}\\
\ac{ROF}                        	   & 0.5                                \\
CP duration                           & 32 samples                         \\
CS duration                            & 16 samples                         \\
Window length                          & 8 samples                          \\
Window type                            & 4th \ac{RC} \cite{gfdm_5th_cel_network}             \\ \bottomrule
\end{tabular}
\end{table}

\begin{figure}[!ht]
	\centering
    \captionsetup{justification=centering}
	\scalebox{0.6}{\input{64QAM_AWGN_SISO_UNCODED.tex}}	
	\caption{Performance of the SISO GFDM transceiver operating with uncoded 64-QAM in AWGN channel.}	
	\label{fig:64QAM_SISO_UNCODED}
\end{figure}
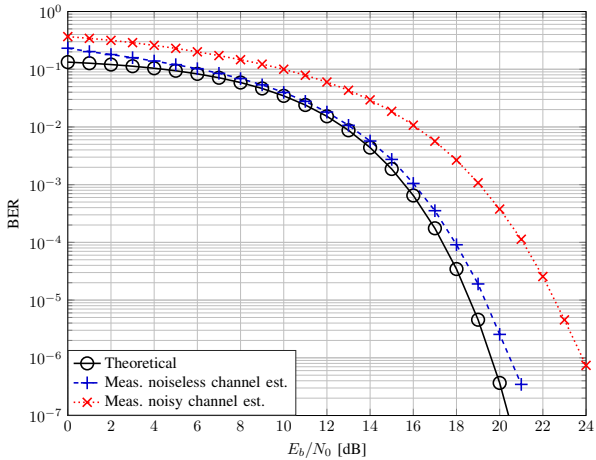

\begin{figure}[!ht]
	\centering
    \captionsetup{justification=centering}
	\scalebox{0.6}{\input{64QAM_AWGN_MIMO_UNCODED.tex}}	
	\caption{Performance of the MIMO GFDM transceiver operating with uncoded 64-QAM in AWGN channel.}	
	\label{fig:64QAM_MIMO_UNCODED}
\end{figure}
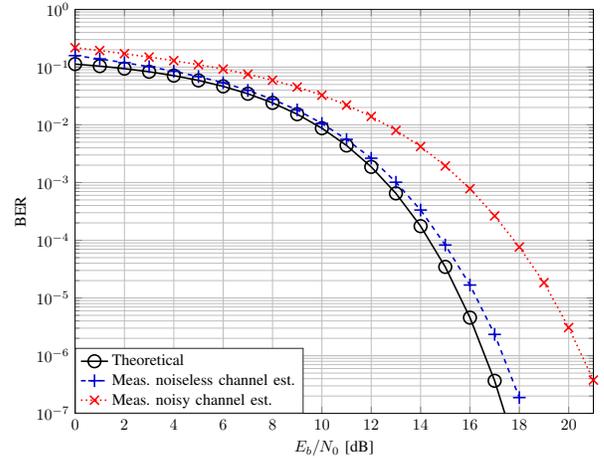

Figure \ref{fig:64QAM_SISO_UNCODED} shows the \ac{BER} performance of the uncoded \ac{SISO} \ac{GFDM} in \ac{AWGN}. The performance of the hardware prototype is shown assuming noiseless channel estimation and noisy channel estimation. It is possible to notice a degradation of approximately 2.5 dB at $\text{BER}=10^{-2}$ caused by the imperfect \ac{CSI} used by the equalizer. However, the performance of the implemented hardware under noiseless channel estimation is close to the theoretical curve, showing that the impairments introduced by coefficient and samples quantizations and by the RF front-ends play a small role in the system performance. 
Figure \ref{fig:64QAM_MIMO_UNCODED} shows the \ac{BER} performance of the uncoded \ac{MIMO} \ac{GFDM} in \ac{AWGN}. As expected, the theoretical curve is 3 dB shifted from the \ac{SISO} case, since half of the transmit power is employed by each transmit antenna \cite{TRSTC-GFDM}. The behavior of the prototype BER curves under noisy or noiseless channel estimation is the same as in the \ac{SISO} case.

The polar code used in the evaluation of the prototype is described in Table \ref{tab:polar_param}.
A simulation comprising the waveform and polar code with the same parameters implemented in hardware was performed and the simulated \ac{BER} curve is used as reference for the measured results. 
\begin{table}[!ht]
\centering
\caption{Polar code parameters}
\label{tab:polar_param}
\begin{tabular}{@{}ll@{}}
\toprule
\multicolumn{1}{c}{\textbf{Parameter}} & \multicolumn{1}{c}{\textbf{Value}} \\ \midrule
Code length (N)                 & 2048                               \\
Shortened bits                  & 32                               \\
Code rate                    & 3/4                                \\
Decoded type                           & \ac{SCD}                           \\ \bottomrule
\end{tabular}
\end{table}

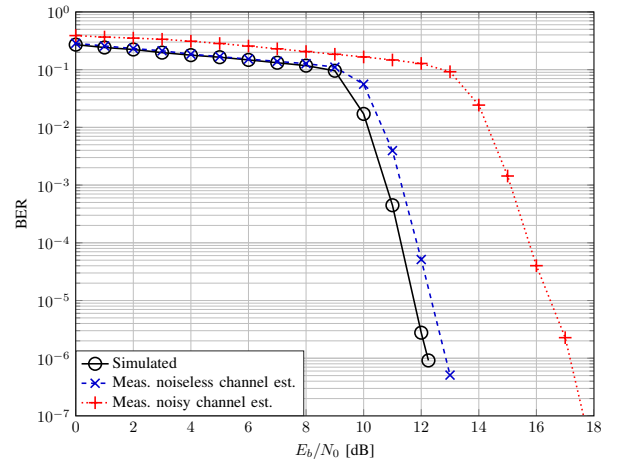
\begin{figure}[!ht]
	\centering
    \captionsetup{justification=centering}
	\scalebox{0.6}{\input{64QAM_AWGN_SISO_CODED.tex}}	
	\caption{Coded SISO-GFDM BER performance under AWGN channel.}	
	\label{fig:64QAM_SISO_CODED}
\end{figure}

\begin{figure}[!ht]
	\centering
    \captionsetup{justification=centering}
	\scalebox{0.6}{\input{64QAM_AWGN_MIMO_CODED.tex}}	
	\caption{Coded MIMO-GFDM BER performance under AWGN channel.}	
	\label{fig:64QAM_MIMO_CODED}
\end{figure}
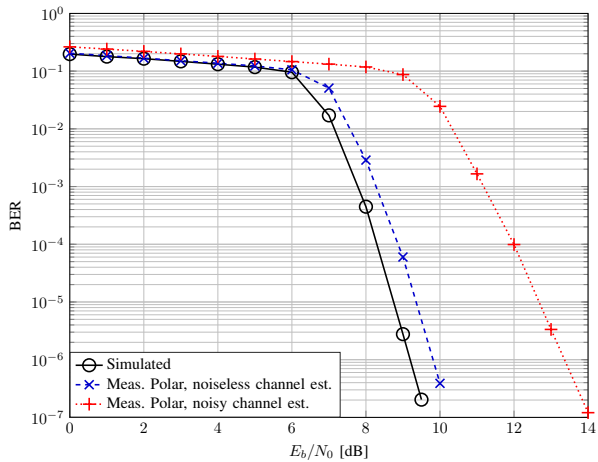

Figure \ref{fig:64QAM_SISO_CODED} shows the \ac{BER} performance of the coded \ac{SISO} \ac{GFDM} under \ac{AWGN}. The evaluation was also performed under noiseless and noisy channel estimations.
Figure \ref{fig:64QAM_MIMO_CODED} shows the \ac{BER} performance of the coded \ac{MIMO} \ac{GFDM} also under \ac{AWGN} channel.
For both, \ac{SISO} and \ac{MIMO} test cases, the noisy channel estimation caused a degradation of approximately 3.6 dB at $\text{BER}=10^{-6}$. For noiseless channel estimation, the performance is within 1 dB from the simulated error rate. Again, the performance loss introduced by practical implementation issues can be considered satisfactory. 

\subsection{\ac{OOB} emissions analysis}
\begin{figure}[!ht]
	\centering
    \captionsetup{justification=centering}
	\includegraphics[width=0.45\textwidth]{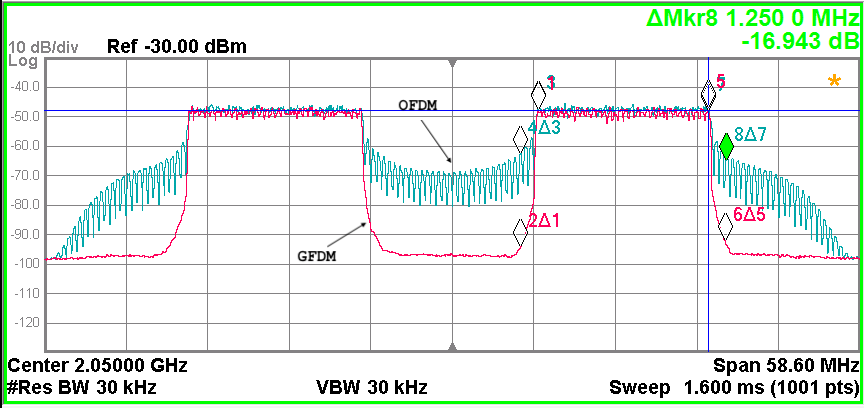}
    \caption{GFDM and OFDM OOB emissions.}
    \label{fig:OOB_GFDM_OFDM}
\end{figure}

\ac{OOB} emission is an important \ac{KPI} for remote areas applications because it is likely that \ac{CR} technologies will be used to exploit \ac{TVWS} and, therefore, to reduce cost \cite{future_CR}. Figure \ref{fig:OOB_GFDM_OFDM} compares the \ac{OOB} emission from \ac{GFDM} and \ac{OFDM} signals. 

Both signals spans 36 MHz in total and the central 12 MHz portion are switched off. The \ac{OFDM} signal presents very high \ac{OOB} emissions, achieving -20 dBc at the center of the unoccupied channel, which is almost 30 dB higher than the noise floor. These emissions can hinder the use of \ac{OFDM} in applications where coexistence with other legacy technologies and spectrum mobility are required. \ac{GFDM}, on the other hand, presents very low \ac{OOB} emissions, achieving almost -50 dBc at the center of the unoccupied channel. In fact, the GFDM \ac{OOB} emissions cannot be distinguished from the measurement equipment error floor. \ac{GFDM} can be used in \ac{TVWS} scenarios where incumbents users must be protected against harmful interference.

%% file: 64QAM_AWGN_SISO_UNCODED.tex
%
\definecolor{mycolor0}{rgb}{0.00,0.00,0.00}%
\definecolor{mycolor1}{rgb}{0.00,0.00,0.80}%
\definecolor{mycolor2}{rgb}{1.00,0.00,0.00}%
\begin{tikzpicture}

\begin{axis}[%
width=4.521in,
height=3.527in,
at={(0.758in,0.519in)},
scale only axis,
xmajorgrids,
yminorticks=true,
ymajorgrids,
yminorgrids,
axis background/.style={fill=white},
legend style={legend cell align=left,align=left,draw=white!15!black,anchor=south west,at={(0,0)}},
xmin=0.000000,
xmax=24.000000,
xlabel={$E_b/N_0$ [dB]},
ymin=0.0000001,
ymax=1.000000,
ylabel={BER},
ymode=log]

\addplot [color=mycolor0,solid,line width=1.0pt,mark size=4.0, mark=o, mark options={solid}]
table[row sep=crcr]{%
	0.000000000000	0.132835614246\\
	1.000000000000	0.127186083458\\
	2.000000000000	0.120567927337\\
	3.000000000000	0.112884060744\\
	4.000000000000	0.104068959693\\
	5.000000000000	0.094113491129\\
	6.000000000000	0.083095705356\\
	7.000000000000	0.071213788457\\
	8.000000000000	0.058812764964\\
	9.000000000000	0.046391152619\\
	10.000000000000	0.034570355714\\
	11.000000000000	0.024013197685\\
	12.000000000000	0.015294130637\\
	13.000000000000	0.008752517703\\
	14.000000000000	0.004389205102\\
	15.000000000000	0.001869322901\\
	16.000000000000	0.000649968549\\
	17.000000000000	0.000175503455\\
	18.000000000000	0.000034536526\\
	19.000000000000	0.000004569921\\
	20.000000000000	0.000000367276\\
	21.000000000000	0.000000015769\\
};
\addlegendentry{Theoretical};

\addplot [color=mycolor1,dashed,line width=1.0pt,mark size=4.0, mark=+, mark options={solid}]
table[row sep=crcr]{%
	0.000000000000	0.231151516945\\
	1.000000000000	0.202188153815\\
	2.000000000000	0.179724738240\\
	3.000000000000	0.158643818176\\
	4.000000000000	0.138774844700\\
	5.000000000000	0.119980782834\\
	6.000000000000	0.102220518330\\
	7.000000000000	0.085020757084\\
	8.000000000000	0.068850896801\\
	9.000000000000	0.053631311574\\
	10.000000000000	0.039662276739\\
	11.000000000000	0.027734098689\\
	12.000000000000	0.018106490293\\
	13.000000000000	0.010662860563\\
	14.000000000000	0.005702700808\\
	15.000000000000	0.002739474891\\
	16.000000000000	0.001050483002\\
	17.000000000000	0.000354124316\\
	18.000000000000	0.000091052531\\
	19.000000000000	0.000019051974\\
	20.000000000000	0.000002545058\\
	21.000000000000	0.000000346286\\
};
\addlegendentry{Meas. noiseless channel est.};

\addplot [color=mycolor2,dotted,line width=1.0pt,mark size=4.0, mark=x, mark options={solid}]
table[row sep=crcr]{%
	0.000000000000	0.367270387115\\
	1.000000000000	0.342878999767\\
	2.000000000000	0.316464428403\\
	3.000000000000	0.288787941293\\
	4.000000000000	0.259436520160\\
	5.000000000000	0.230043633837\\
	6.000000000000	0.200385065866\\
	7.000000000000	0.172361950403\\
	8.000000000000	0.146240406113\\
	9.000000000000	0.122450651610\\
	10.000000000000	0.099995350590\\
	11.000000000000	0.078689531812\\
	12.000000000000	0.060021010326\\
	13.000000000000	0.043239619612\\
	14.000000000000	0.029445971988\\
	15.000000000000	0.018662780369\\
	16.000000000000	0.010719455817\\
	17.000000000000	0.005662491986\\
	18.000000000000	0.002662138411\\
	19.000000000000	0.001078669386\\
	20.000000000000	0.000377313849\\
	21.000000000000	0.000113279880\\
	22.000000000000	0.000025600590\\
	23.000000000000	0.000004535341\\
	24.000000000000	0.000000739801\\
};
\addlegendentry{Meas. noisy channel est.};
\end{axis}
\end{tikzpicture}%

%% file: 64QAM_AWGN_MIMO_UNCODED.tex
%
\definecolor{mycolor0}{rgb}{0.00,0.00,0.00}%
\definecolor{mycolor1}{rgb}{0.00,0.00,0.80}%
\definecolor{mycolor2}{rgb}{1.00,0.00,0.00}%
\begin{tikzpicture}

\begin{axis}[%
width=4.521in,
height=3.527in,
at={(0.758in,0.519in)},
scale only axis,
xmajorgrids,
yminorticks=true,
ymajorgrids,
yminorgrids,
axis background/.style={fill=white},
legend style={legend cell align=left,align=left,draw=white!15!black,anchor=south west,at={(0,0)}},
xmin=0.000000,
xmax=21.000000,
xlabel={$E_b/N_0$ [dB]},
ymin=0.0000001,
ymax=1.000000,
ylabel={BER},
ymode=log]

\addplot [color=mycolor0,solid,line width=1.0pt,mark size=4.0, mark=o, mark options={solid}]
table[row sep=crcr]{%
	0.000000000000	0.112884060744\\
	1.000000000000	0.104068959693\\
	2.000000000000	0.094113491129\\
	3.000000000000	0.083095705356\\
	4.000000000000	0.071213788457\\
	5.000000000000	0.058812764964\\
	6.000000000000	0.046391152619\\
	7.000000000000	0.034570355714\\
	8.000000000000	0.024013197685\\
	9.000000000000	0.015294130637\\
	10.000000000000	0.008752517703\\
	11.000000000000	0.004389205102\\
	12.000000000000	0.001869322901\\
	13.000000000000	0.000649968549\\
	14.000000000000	0.000175503455\\
	15.000000000000	0.000034536526\\
	16.000000000000	0.000004569921\\
	17.000000000000	0.000000367276\\
	18.000000000000	0.000000015769\\
	19.000000000000	0.000000000308\\
	20.000000000000	0.000000000002\\
	21.000000000000	0.000000000000\\
};
\addlegendentry{Theoretical};

\addplot [color=mycolor1,dashed,line width=1.0pt,mark size=4.0, mark=+, mark options={solid}]
table[row sep=crcr]{%
0.000000000000	0.158545417762\\
1.000000000000	0.138478713670\\
2.000000000000	0.119472096199\\
3.000000000000	0.101836149897\\
4.000000000000	0.084759723385\\
5.000000000000	0.068568678713\\
6.000000000000	0.053646570081\\
7.000000000000	0.039564730644\\
8.000000000000	0.027656762209\\
9.000000000000	0.018125331879\\
10.000000000000	0.010580036497\\
11.000000000000	0.005588669521\\
12.000000000000	0.002638135323\\
13.000000000000	0.001011674913\\
14.000000000000	0.000334084580\\
15.000000000000	0.000082724929\\
16.000000000000	0.000016770004\\
17.000000000000	0.000002338828\\
18.000000000000	0.000000188090\\
};
\addlegendentry{Meas. noiseless channel est.};

\addplot [color=mycolor2,dotted,line width=1.0pt,mark size=4.0, mark=x, mark options={solid}]
table[row sep=crcr]{%
	0.000000000000	0.217170322397\\
	1.000000000000	0.193697817776\\
	2.000000000000	0.170782178700\\
	3.000000000000	0.149421148453\\
	4.000000000000	0.129407002959\\
	5.000000000000	0.110318743865\\
	6.000000000000	0.092331062347\\
	7.000000000000	0.075273929119\\
	8.000000000000	0.059351797720\\
	9.000000000000	0.045119757415\\
	10.000000000000	0.032578966559\\
	11.000000000000	0.021974337270\\
	12.000000000000	0.013985289046\\
	13.000000000000	0.008008799052\\
	14.000000000000	0.004205561467\\
	15.000000000000	0.001938922918\\
	16.000000000000	0.000781467941\\
	17.000000000000	0.000264343430\\
	18.000000000000	0.000076278546\\
	19.000000000000	0.000018446029\\
	20.000000000000	0.000003075521\\
	21.000000000000	0.000000377912\\
};
\addlegendentry{Meas. noisy channel est.};
\end{axis}
\end{tikzpicture}%

%% file: 64QAM_AWGN_SISO_CODED.tex
%
\definecolor{mycolor0}{rgb}{0.00,0.00,0.00}%
\definecolor{mycolor1}{rgb}{0.00,0.00,0.80}%
\definecolor{mycolor2}{rgb}{1.00,0.00,0.00}%
\begin{tikzpicture}

\begin{axis}[%
width=4.521in,
height=3.527in,
at={(0.758in,0.519in)},
scale only axis,
xmajorgrids,
yminorticks=true,
ymajorgrids,
yminorgrids,
axis background/.style={fill=white},
legend style={legend cell align=left,align=left,draw=white!15!black,anchor=south west,at={(0,0)}},
xmin=0.000000,
xmax=18.000000,
xlabel={$E_b/N_0$ [dB]},
ymin=0.0000001,
ymax=1.000000,
ylabel={BER},
ymode=log]

\addplot [color=mycolor0,solid,line width=1.0pt,mark size=4.0, mark=o, mark options={solid}]
table[row sep=crcr]{%
	0.000000000000	0.271206555974\\
	1.000000000000	0.242011354047\\
	2	0.2236243386243385\\
	3	0.197373393801965\\
	4	0.17838246409675\\
	5	0.164304610733182\\
	6	0.147581254724112\\
	7	0.1317554799697655\\
	8	0.11734693877551\\	
	9	0.0956868858654573\\
	10	0.0171091584782061\\
	11	0.0004475709518458235\\
	12	2.770159758162385e-6\\
	12.25	9.15876872745456e-07\\
};
\addlegendentry{Simulated};

\addplot [color=mycolor1,dashed,line width=1.0pt,mark size=4.0, mark=x,mark options={solid}]
table[row sep=crcr]{%
	0.000000000000	0.286236555974\\
	1.000000000000	0.257211354047\\
	2.000000000000	0.232518024447\\
	3.000000000000	0.212213633563\\
	4.000000000000	0.185574898898\\
	5.000000000000	0.168256519722\\
	6.000000000000	0.152401127007\\
	7.000000000000	0.139021730098\\
	8.000000000000	0.127109537152\\
	9.000000000000	0.110730474923\\
	10.000000000000	0.055925831671\\
	11.000000000000	0.003956757936\\
	12.000000000000	0.000051330259\\
	13.000000000000	0.000000510264\\
};
\addlegendentry{Meas. noiseless channel est.};

\addplot [color=mycolor2,dotted,line width=1.0pt,mark size=4.0, mark=+,mark options={solid}]
table[row sep=crcr]{%
	0.000000000000	0.389442685620\\
	1.000000000000	0.367350666168\\
	2.000000000000	0.351990107889\\
	3.000000000000	0.336196315301\\
	4.000000000000	0.310906715119\\
	5.000000000000	0.283864034541\\
	6.000000000000	0.256288173093\\
	7.000000000000	0.229791978284\\
	8.000000000000	0.205914625709\\
	9.000000000000	0.184892717162\\
	10.000000000000	0.166061869841\\
	11.000000000000	0.147109924658\\
	12.000000000000	0.128046316686\\
	13.000000000000	0.092061829410\\
	14.000000000000	0.024314702094\\
	15.000000000000	0.001436826178\\
	16.000000000000	0.000040163980\\
	17.000000000000	0.000002271606\\
	18.000000000000	0.000000016936\\
};
\addlegendentry{Meas. noisy channel est.};

\end{axis}
\end{tikzpicture}%

%% file: 64QAM_AWGN_MIMO_CODED.tex
%
\definecolor{mycolor0}{rgb}{0.00,0.00,0.00}%
\definecolor{mycolor1}{rgb}{0.00,0.00,0.80}%
\definecolor{mycolor2}{rgb}{1.00,0.00,0.00}%
\begin{tikzpicture}

\begin{axis}[%
width=4.521in,
height=3.527in,
at={(0.758in,0.519in)},
scale only axis,
xmajorgrids,
yminorticks=true,
ymajorgrids,
yminorgrids,
axis background/.style={fill=white},
legend style={legend cell align=left,align=left,draw=white!15!black,anchor=south west,at={(0,0)}},
xmin=0.000000,
xmax=14.000000,
xlabel={$E_b/N_0$ [dB]},
ymin=0.0000001,
ymax=1.00000,
ylabel={BER},
ymode=log]

\addplot [color=mycolor0,solid,line width=1.0pt,mark size=4.0, mark=o, mark options={solid}]
table[row sep=crcr]{%
	0	0.197373393801965\\	
	1	0.178382464096750\\	
	2	0.164304610725\\	
	3	0.147581254724112\\	
	4	0.1317554799697655\\	
	5	0.117346938775510\\	
	6	0.0956868858654573\\	
	7	0.0171091584782061\\	
	8	0.0004475709518458235\\	
	9	2.770159758162385e-6\\
	9.5	2.03152684975123e-7\\		
};
\addlegendentry{Simulated};

\addplot [color=mycolor1,dashed,line width=1.0pt,mark size=4.0, mark=x,mark options={solid}]
table[row sep=crcr]{%
	0.000000000000	0.204145439990\\
	1.000000000000	0.185340077352\\
	2.000000000000	0.167734907346\\
	3.000000000000	0.151388917769\\
	4.000000000000	0.137096673855\\
	5.000000000000	0.123019665836\\
	6.000000000000	0.105410958904\\
	7.000000000000	0.050776186222\\
	8.000000000000	0.002875013850\\
	9.000000000000	0.000060120964\\
	10.000000000000	0.000000389437\\
};
\addlegendentry{Meas. Polar, noiseless channel est.};

\addplot [color=mycolor2,dotted,line width=1.0pt,mark size=4.0, mark=+,mark options={solid}]
table[row sep=crcr]{%
	0.000000000000	0.263325823482\\
	1.000000000000	0.241067409700\\
	2.000000000000	0.219453487890\\
	3.000000000000	0.198754383959\\
	4.000000000000	0.179845732231\\
	5.000000000000	0.162015781674\\
	6.000000000000	0.146318763503\\
	7.000000000000	0.132201124591\\
	8.000000000000	0.118009414500\\
	9.000000000000	0.087113248573\\
	10.000000000000	0.024509289652\\
	11.000000000000	0.001658877216\\
	12.000000000000	0.000098942305\\
	13.000000000000	0.000003344494\\
	14.000000000000	0.000000120810\\
};
\addlegendentry{Meas. Polar, noisy channel est.};

\end{axis}
\end{tikzpicture}%

%% file: 4.tex
\section{Conclusion}\label{sec:4}

\ac{5G} for remote area networks will face several challenges to provide reliable high throughput Internet access in low populated areas. Robustness for covering large distances and low \ac{OOB} emissions are important \acp{KPI} to be achieved. This paper has shown that modern waveforms and channel codes can be used to reduce the undesired emissions and improve \ac{BER} performance under \ac{AWGN} channels. The coded and uncoded \ac{BER} performance of the implemented transceiver, under noiseless channel estimation, is close to the theoretical curve. This result indicates that the digital signal processing chain, as well as the RF analog signal chain, play a small hole in the overall system performance. Conversely, the error rate under noisy channel estimation suggests that there is room for improvement in the estimation process.



The measured \ac{OOB} emissions confirm that \ac{GFDM} has a very well confined spectrum, while \ac{OFDM} spreads harmful interference when it is not filtered. Hence, the former is more suitable for remote areas applications than the later, since \ac{CR} techniques is likely to be employed to allow for \ac{TVWS} exploitation and fragmented spectrum allocation.